\documentclass[a4paper,11pt]{article}
\usepackage{jinstpub} % for details on the use of the package, please see the JINST-author-manual
\usepackage{lineno}
\usepackage{amsmath}

\title{\boldmath Multi-threaded Simulation Software for the JUNO Experiment}

\author[a,b]{P.D.~Yu}
\author[a,1]{T.~Lin\note{Corresponding author.}}
\author[a,2]{Z.Y.~Deng\note{Corresponding author.}}
\author[a,b]{G.F.~Cao}
\author[a,b]{Y.X.~Hu}
\author[a]{S.C.~Blyth}
\author[a]{J.H.~Zou}
\author[a,b]{W.D.~Li}
\author[c]{X.T.~Huang}
\author[c]{T.~Li}
\author[a,b]{Y.~Peng}

\affiliation[a]{Institute of High Energy Physics, Chinese Academy of Sciences, Beijing 100049, China}
\affiliation[b]{University of Chinese Academy of Sciences, Beijing 100049, China}
\affiliation[c]{Shandong University, Qingdao 266237, China}

\emailAdd{lintao@ihep.ac.cn}
\emailAdd{dengzy@ihep.ac.cn}

\abstract{The Jiangmen Underground Neutrino Observatory (JUNO) is a multi-purpose experiment under construction in southern China. JUNO aims to determine the neutrino mass ordering and precisely measure the neutrino oscillation parameters by detecting reactor neutrinos from nuclear power plants. In addition to reactor neutrinos, JUNO can study atmospheric neutrinos, solar neutrinos, geo-neutrinos, supernova burst neutrinos, the Diffuse Supernova Neutrino Background (DSNB), and nucleon decay, covering an energy range from keV to TeV. JUNO consists of a 20 kton liquid scintillator detector, equipped with 17,612 20-inch PMTs and 25,600 3-inch PMTs, achieving a 3\% energy resolution (at 1~MeV). Precise simulation plays a crucial role in the experiment. Developed using the SNiPER framework and Geant4 toolkit, the JUNO simulation software is a key component of the JUNO offline software (JUNOSW). The large detector size and the broad energy range of interest pose challenges for detector simulation in terms of CPU time and memory consumption. As the computing nodes shift to include multiple integrated CPU cores, traditional single-threaded computing model can lead to large memory footprints and inefficient use of system resources. Multi-threading on many-core architectures can significantly improve resource utilization. This paper introduces the design and implementation of multi-threaded simulation software for the JUNO experiment.}


\begin{document}
\maketitle
\flushbottom

%%%%%%%%%%%%%%%%%%%%%%%%%%%%%%%%%%%%%%%%%%%%%%%%%%%%%
% main text
%%%%%%%%%%%%%%%%%%%%%%%%%%%%%%%%%%%%%%%%%%%%%%%%%%%%%

\section{Introduction} 
\label{sec:introduction}   

\begin{figure}[htb]\centering
\includegraphics[width=85mm]{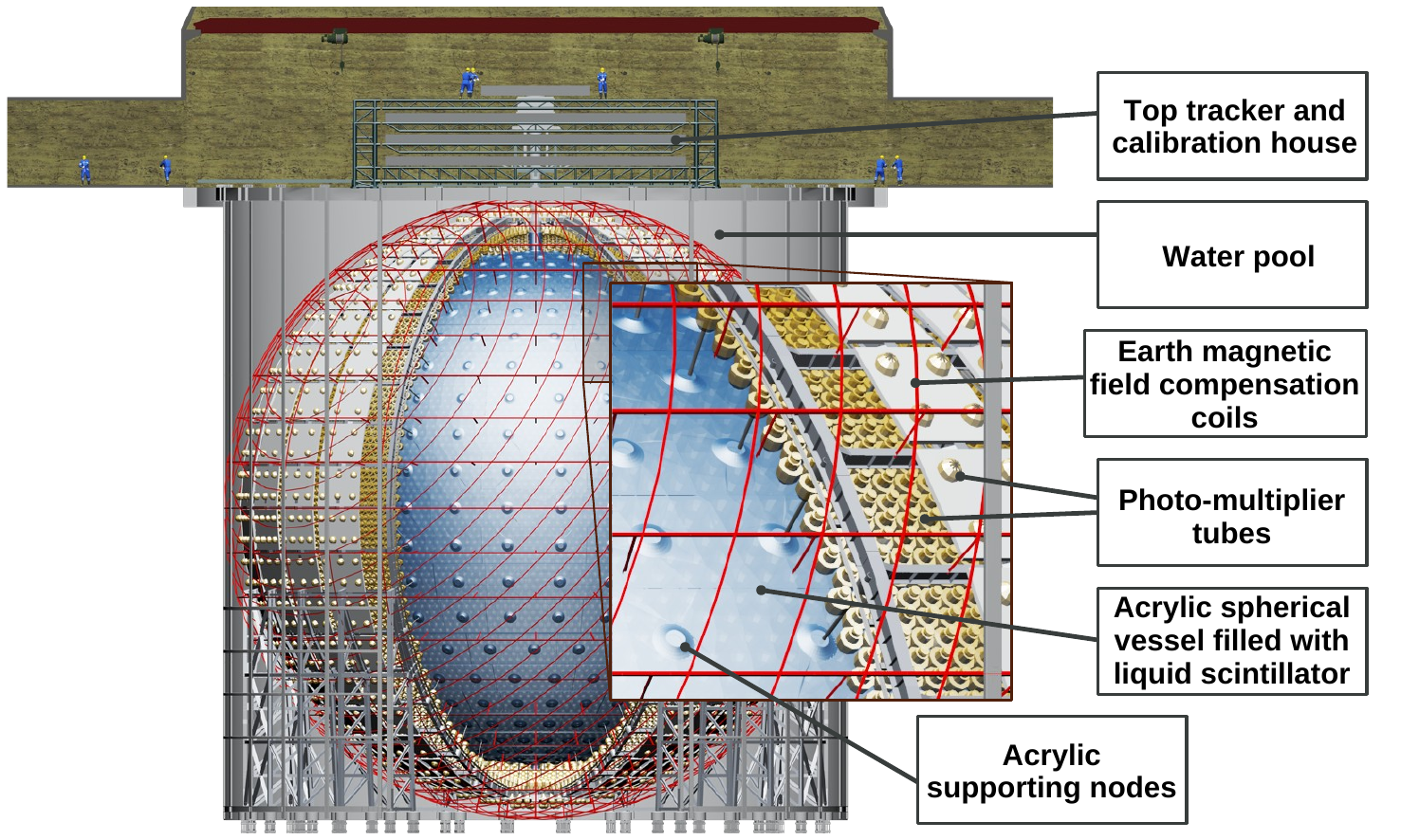}
\caption{JUNO Detector}
\label{fig:Detector}
\end{figure}

The Jiangmen Underground Neutrino Observotary (JUNO)~\cite{JUNO:2015zny,JUNO:2015sjr,JUNO:2021vlw} is a multi-purpose experiment under construction in southern China, expected to finish the construction and start the detector filling by the end of 2024. JUNO is designed to determine the neutrino mass ordering and precisely measure the neutrino oscillation parameters by detecting reactor neutrinos from the Yangjiang and Taishan nuclear power plants. JUNO is also able to observe supernova neutrinos, study the atmospheric neutrinos, solar neutrinos and geo-neutrinos, and perform exotic searches. The JUNO detector is designed with a 20~kton liquid scintillator detector of unprecedented 3{\%} energy resolution (at 1~MeV) at 700~meter deep underground. As shown in Figure \ref{fig:Detector}, a central detector (CD) contains an acrylic sphere, in which the liquid scintillator is located, surrounded by 17,612 20-inch PMTs and 25,600 3-inch PMTs to detect the scintillation and Cherenkov light with about 78{\%} photocathode coverage. The central detector is immersed in a water pool (WP) instrumented with 2,400 20-inch PMTs to serve as a water Cherenkov detector. The top tracker (TT) is placed on top of the water pool. The water Cherenkov detector and the top tracker form the veto system of JUNO to track and veto muons crossing the detector.

\begin{figure}[htb]\centering
\includegraphics[width=85mm]{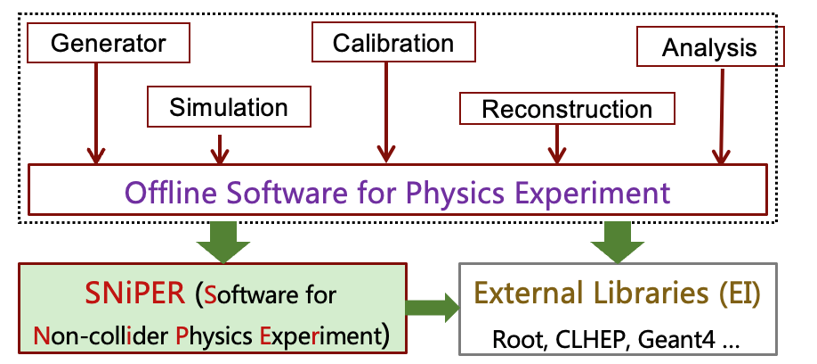}
\caption{JUNO offline software components}
\label{fig:JUNOSW}
\end{figure}

Offline software is essential for each particle physics experiment in the entire lifecycle of the experiment, from the experiment design, detector construction, to the data analysis. As shown in Figure \ref{fig:JUNOSW}, the JUNO offline software (JUNOSW) is developed based on the SNiPER~\cite{Zou:2015ioy} framework, and some other external libraries, such as ROOT~\cite{Brun:1997pa}, Geant4~\cite{Allison:2006ve, Allison:2016lfl}, and CLHEP~\cite{Lonnblad:1994np}. The SNiPER framework is a general purpose light-weighted software framework, developed with bi-language C++ and Python. It was originally developed for JUNO, but also adopted by other experiments for example STCF~\cite{Huang:2022bkz}, HERD~\cite{Cagnoli:2024hgd}, and nEXO~\cite{Adhikari_2022}. 
% SNiPER also provides interfaces to support multi-threaded simulation, reconstruction and analysis. 

%

JUNO simulation software is a key component of JUNOSW. The simulation software is developed based on Geant4 toolkit, with the simulation event loop fully integrated with the SNiPER. Since rich physics topics can be studied with JUNO, the simulation software can be used to generate different physics events with the energy range from keV to TeV. Full simulation of the primary particles and all the secondary optical photons in JUNO detector is implemented to meet the requirements on detector simulation precision. The simulation software will be introduced in section \ref{sec:simulation}. Due to the large detector size of JUNO, high energy events can produce tens of million optical photons in the central detector, which pose challenges to detector simulation both in CPU time and memory consumption. Traditional massive Monte Carlo data production is done by having many independent single-threaded jobs running on many CPU cores, where each job runs independently on a single core. However, as CPU clock rates have plateaued, computing nodes have changed to including multiple integrated CPU cores. Traditional single-threaded computing model can lead to large memory footprints and inefficient use of system resources. The memory footprint for JUNO detector simulation jobs will be introduced in section \ref{sec:memory}. Multi-threading on many-core architecture can significantly improve utilization of computing resources. The implementation of multi-threaded simulation in JUNOSW will be introduced in section \ref{sec:multithread}, and the performance results will be introduced in section \ref{sec:performance}. Section \ref{sec:summary} is the summary. 

\section{JUNO simulation software}
\label{sec:simulation}
\begin{figure}[htb]\centering
\includegraphics[width=85mm]{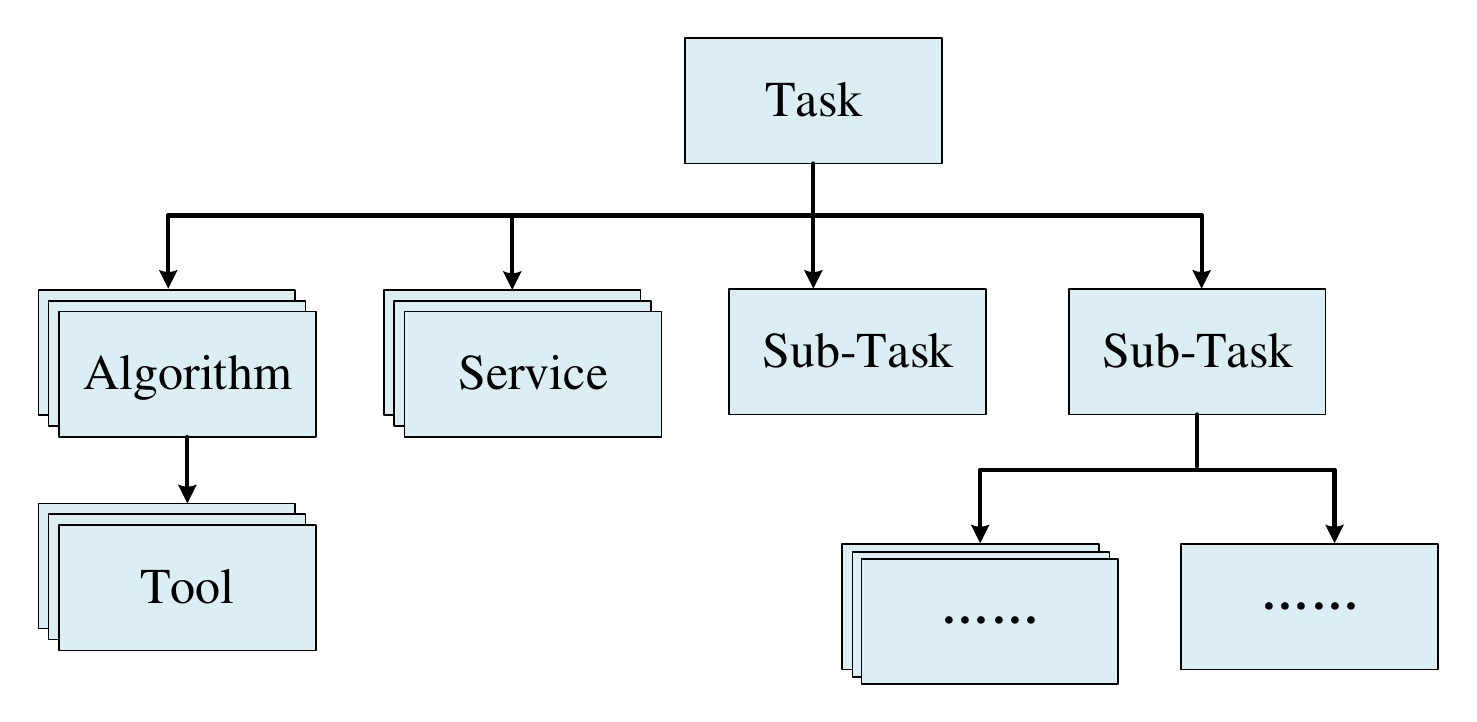}
\caption{SNiPER components}
\label{fig:SNiPER}
\end{figure}

\begin{figure}[htb]\centering
\includegraphics[width=85mm]{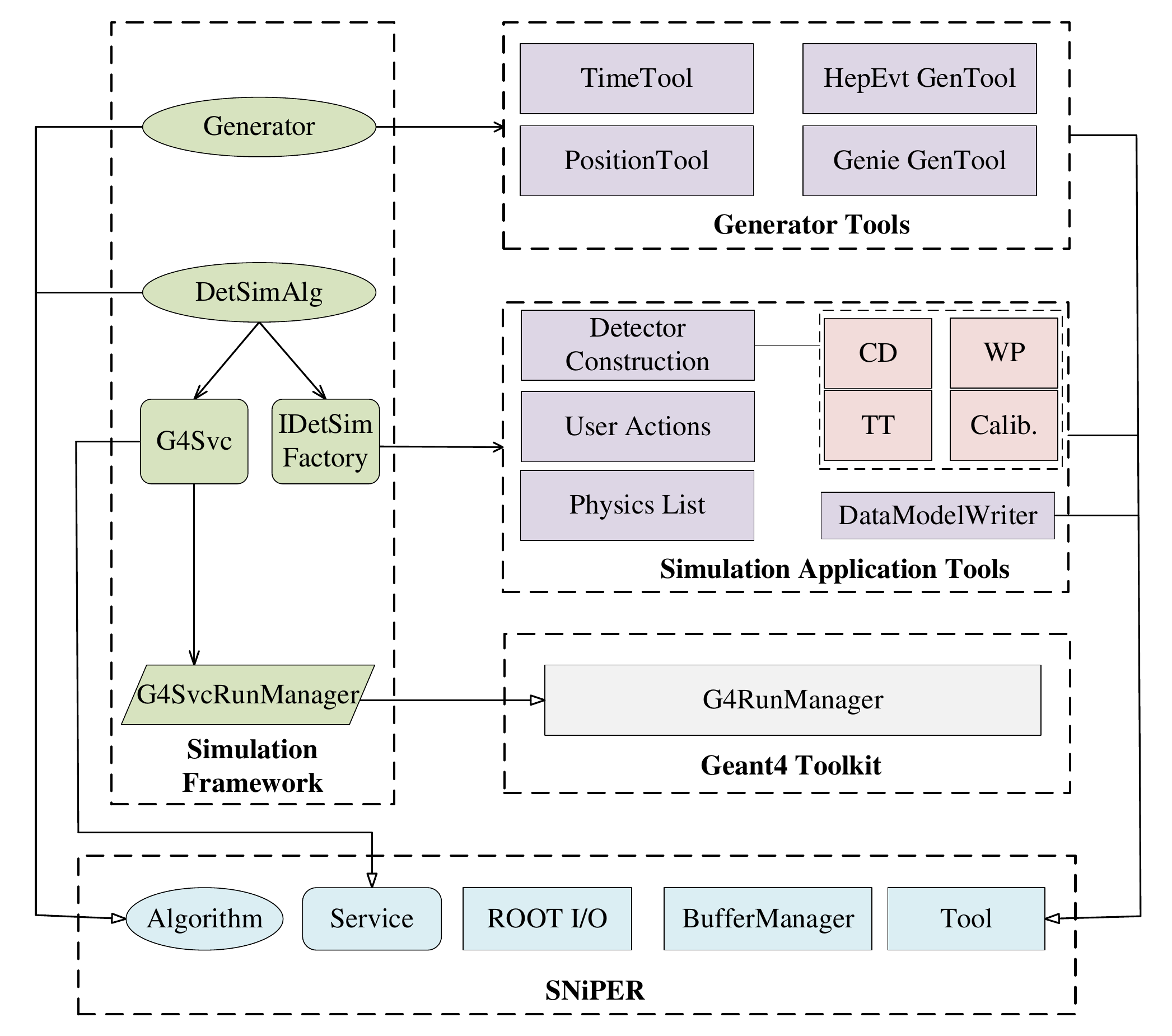}
\caption{Detector simulation software}
\label{fig:DetSim}
\end{figure}

JUNO simulation software~\cite{Lin:2022htc} is developed based on Geant4, fully taking advantage of elements provided by the SNiPER framework. Figure \ref{fig:SNiPER} shows the main components in the SNiPER framework, which consists of \texttt{Task}, \texttt{Algorithm}, \texttt{Tool} and \texttt{Service} components. A \texttt{Task} is an application manager to assemble and control the other components. An \texttt{Algorithm} is a basic unit in data processing. A \texttt{Tool} is a lightweight routine that can be invoked within an \texttt{Algorithm}. A \texttt{Service} is a unit for common functions that can be called anywhere when necessary. As shown in Figure \ref{fig:DetSim}, a simulation framework is implemented with a generator algorithm and a detector simulation algorithm in a task. The generator algorithm calls various generator tools to integrate the different physics generators into the simulation chain. The available generators consist of cosmic muons, reactor neutrinos, atmospheric neutrinos, solar neutrinos, supernova burst neutrinos, DSNB, natural radioactivities, and calibration sources. The detector simulation algorithm invokes a customized Geant4 Run Manager to integrate the underlying Geant4 kernel into the SNiPER. A service named \texttt{G4Svc} manages the run manager, which is invoked during the initialization of the detector simulation algorithm. To configure the geometry, physics list and user action before the initialization of the Geant4 run manager, they are all implemented as tools.


\section{Memory footprint and optimization in simulation}
\label{sec:memory}

Since 2012, there's no obvious increase in CPU clock speeds, but there is a significant increase in the number of processor cores. The memory prices have not decreased significantly, so the ratio of memory to cores tends to decrease. It is now difficult to fully utilize all CPU cores by running multiple jobs on a single computing node, so it is necessary to reduce the amount of memory per core. For high energy events such as muons and atmospheric neutrinos in JUNO simulation, the minimum memory consumption typically exceeds 3~GB. Due to the configuration of computing centers, which provides a fixed amount of memory per CPU core, jobs requiring more memory upon submission will result in the allocation of multiple CPU cores by the computing center to meet the memory requirements. However, only one CPU core is actually utilized to run the job, leaving other cores idle, which results in wasted computational resources and impacts data production for high-energy events. This implies that in large-scale data production for high-energy events, the more jobs submitted to computing centers, the more likely CPU cores will be idle due to memory constraints.            

To address this inefficiency, it is crucial to explore memory optimization techniques that can effectively reduce the memory footprint of simulations while maintaining the accuracy and reliability of the results.

As the generator and detector simulation algorithms always run in the same job, the major memory footprint in the simulation typically comprises generator parameters, detector geometry and material properties, physics processes, hit objects and ROOT I/O. The memory consumption can vary for different event types and energy ranges. There are corresponding optimizations to reduce memory usage. 

The memory usage of generator parameters depends on the types of physics generators. For example, the particle gun tool does not depend on any extra information, so the memory usage is not critical. For some high energy neutrino generators, they need to load the parameters, such as cross sections. For example, the GENIE~\cite{Andreopoulos:2009rq} generator needs to load a file in XML format~\cite{bray1997extensible}, which could be larger than 1 GB. In order to reduce memory usage, the physics generator developers provide a compact XML file with only necessary information. 

The memory usage for detector geometry, material properties, and physics processes takes more than 2 GB.  Even though the JUNO detector structure is not complicated, there are a lot of replicated volumes. The Geant4 smart voxels technique occupies additional memory to optimize the search for intersections of these volumes. Since a precise simulation of the low energy part is important, the high precision physics process \texttt{G4IonPhysicsPHP} is used in JUNO simulation, which loads more than 1 GB of data into memory, tested with Geant4 version 10.4. 

In the JUNO experiment, the core detection technology is the liquid scintillator detector, making optical photon-related physical processes particularly important. Both scintillation and Cherenkov processes can produce photons; the former emits photons isotropically, while the latter emits photons at a specific angle relative to the direction of the primary particle. PMTs serve as the sensitive detectors. When photons are transported to the PMTs, the sensitive detector collects the PMT hits. For the high energy events, millions of PMT hit objects will be created. A PMT hit type contains a PMT ID, a hit time, number of photoelectrons, a local position (relative to the position of PMT), a global position (position in the global coordinate system), a momentum, a polarization and other truth information. To reduce the size of the hit type, a compact version is created that includes only the necessary information: the PMT ID, hit time, hit count, and local position. The object size is reduced from 248 bytes to 40 bytes. Another optimization is to reduce the number of hit objects. Since the sampling period of the PMT will not exceed 1~ns, hits occurring within the same time window (1~ns) are merged into one hit.

\begin{figure}[htb]\centering
\includegraphics[width=65mm]{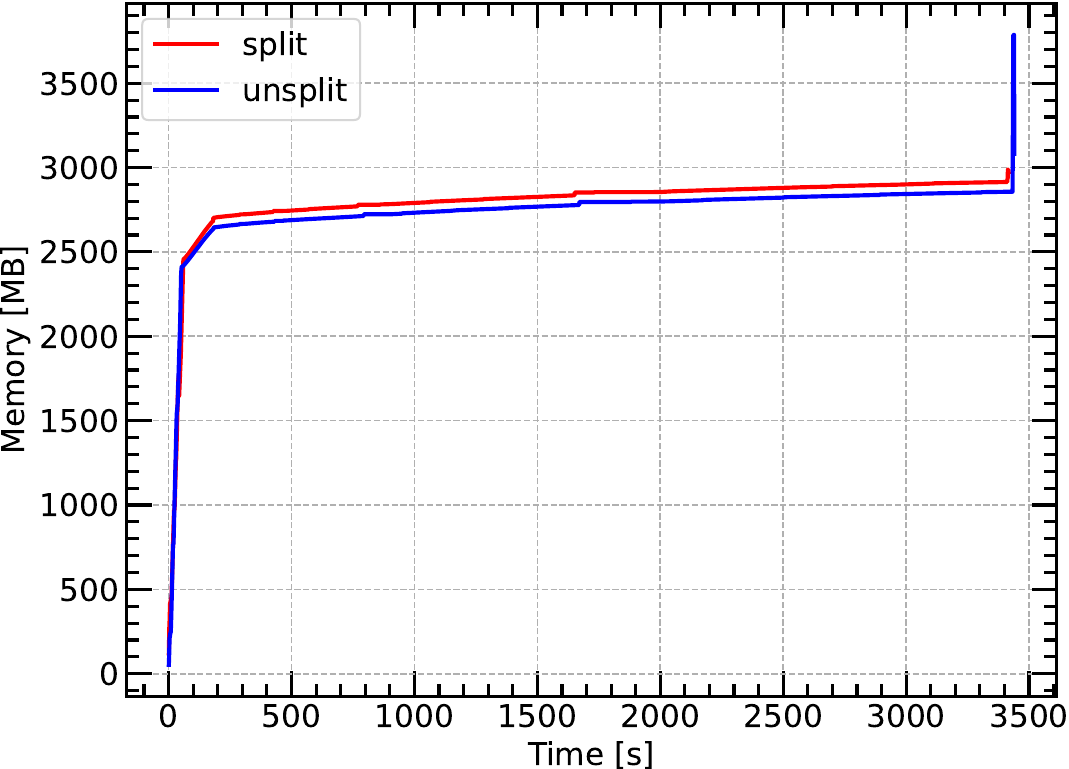}
\caption{Comparison between split and unsplit methods}
\label{fig:memory_split_vs_unsplit}
\end{figure}

\begin{figure}[htb]\centering
\includegraphics[width=85mm]{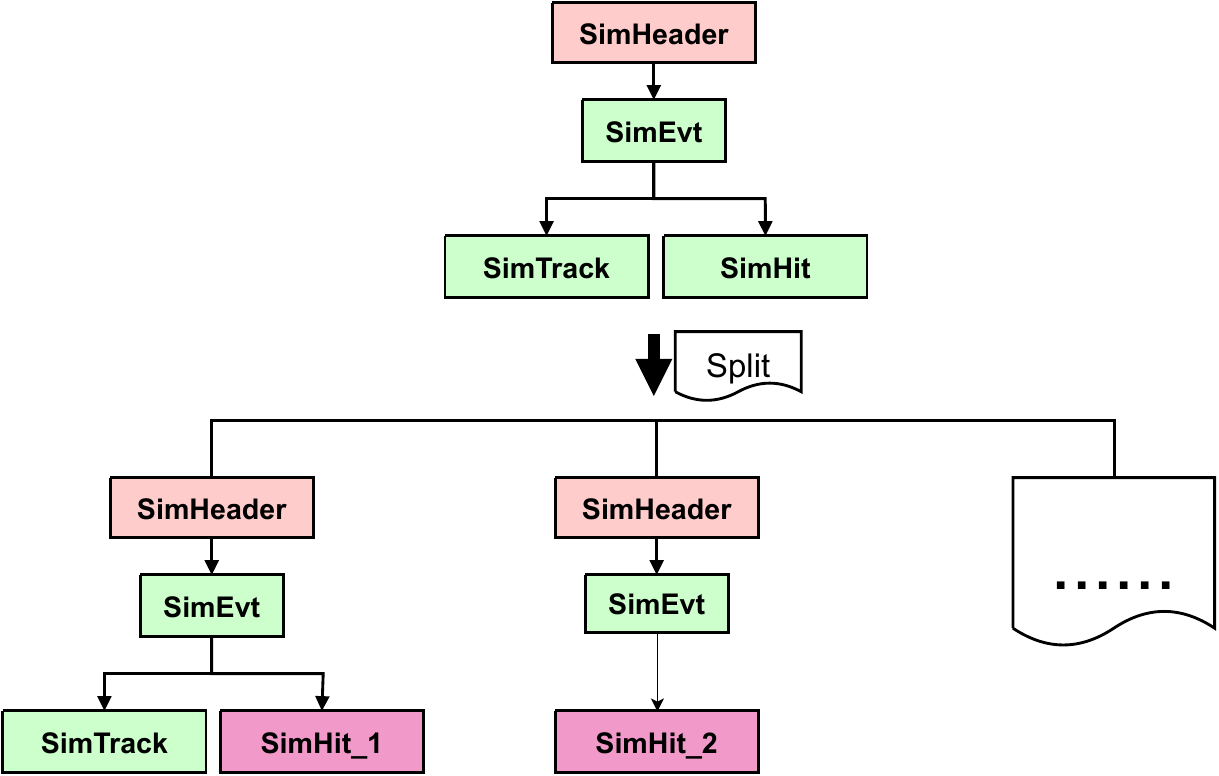}
\caption{The structure of data model spilt.In the sub-events, the first one contains the SimTrack information. After that, only the SimHit information will be split.}
\label{fig:split_event}
\end{figure}

The ROOT I/O also consumes some memory due to the large number of hit objects. As shown in Figure \ref{fig:memory_split_vs_unsplit}, there is a significant spike in memory usage mainly caused by the I/O stage at the end. Because when ROOT writes the hit objects into file, it requires an additional 2-3 times larger memory space to compress the data to ensure a considerable compression ratio. The ``\texttt{split}'' method was developed to solve this problem. As shown in Figure \ref{fig:split_event}, this method can split a collection of hits within an event into multiple parts to complete the output. A Sub-Task is registered in the main task and the \texttt{Incident} mechanism in the SNiPER is used to split an event that generates a large number of hits. In an event generating a large number of hits, when the user-specified hit count is reached, the incident mechanism is used to wake up I/O in subtask for one output until all hits are outputted, thereby reducing memory usage. As shown by the red line in the Figure \ref{fig:memory_split_vs_unsplit}, the situation of a sharp increase in memory usage at the end of program execution has disappeared.%

\begin{figure}[htb]\centering
\includegraphics[width=65mm]{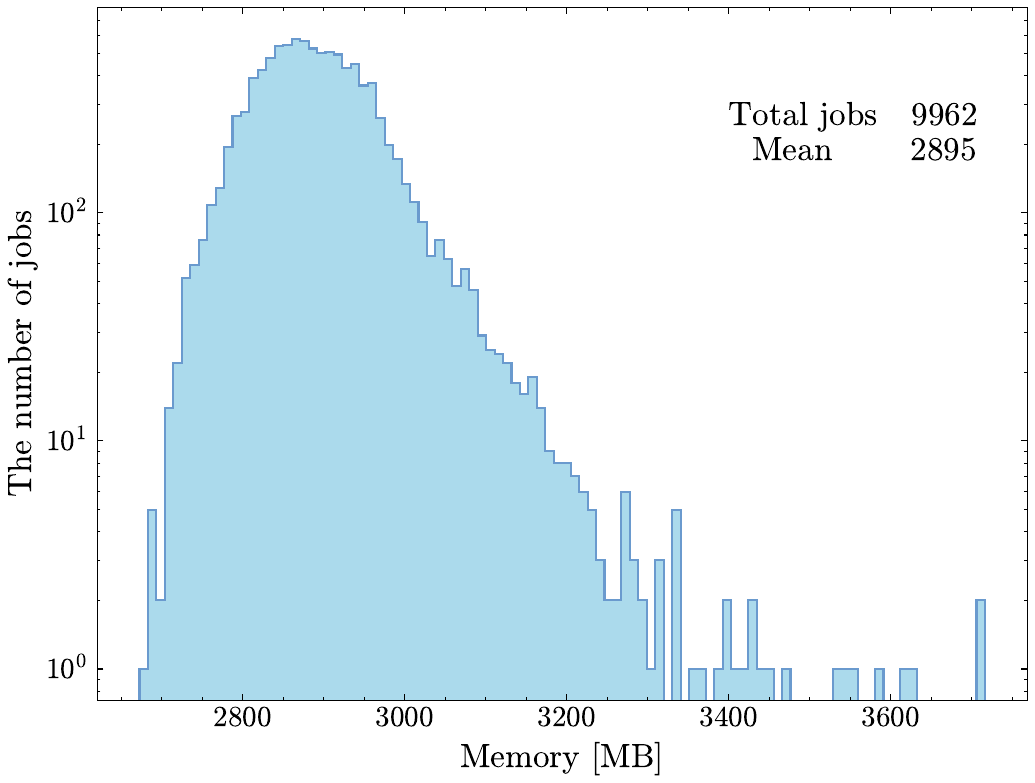}
\caption{Maximum memory of muon simulation jobs}
\label{fig:the_muon_job}
\end{figure}

\begin{figure}[htb]\centering
\includegraphics[width=85mm]{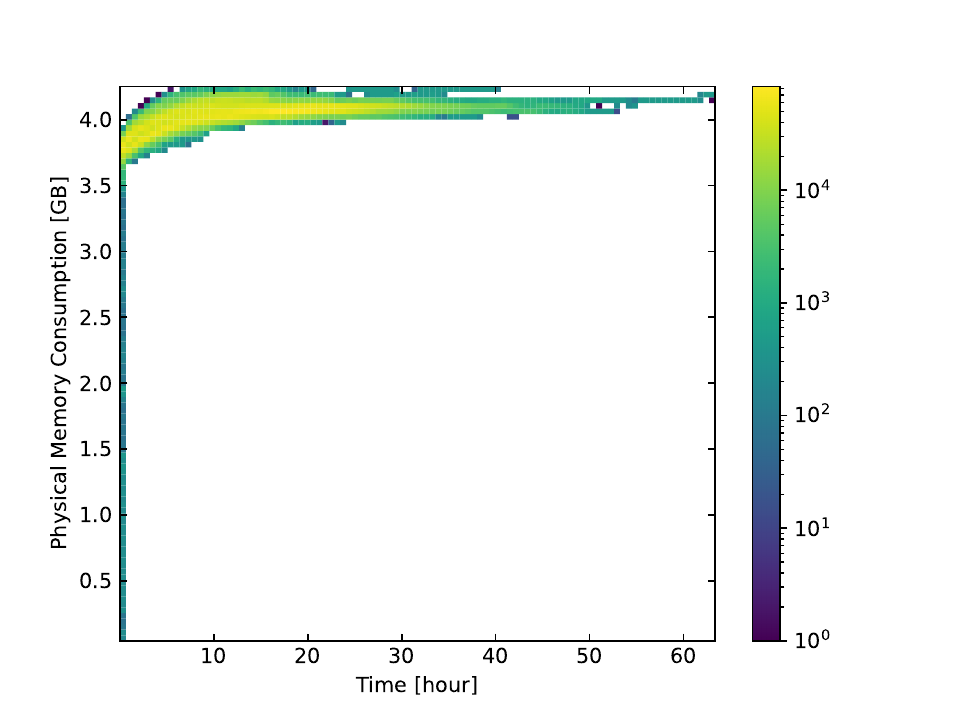}
\caption{Memory consumption versus job running time for atmospheric neutrino simulation}
\label{fig:AtmMem}
\end{figure}

 After optimization, the simulation of a large number of cosmic ray muon events is displayed in Figure \ref{fig:the_muon_job}, with an average memory around 2.9 GB, indicating that the memory bottleneck for high-energy events has been further addressed. Figure \ref{fig:AtmMem} shows the memory consumption for atmospheric neutrino event simulation jobs, with a maximum memory usage reaching 4.5 GB. This is due to the large generator parameters. The memory consumption still can not match with the configuration of computing centers.%

% todo
%Although memory in single-threaded mode are optimized, memory problems still exist in high-energy event %simulation. Due to the difficulty of future core memory and computing resource utilization, JUNO %simulation must maximize memory savings to face with the production of large amount of high energy sample %data. Currently, the memory usage in the physics list and detector geometry construction parts is not %optimized. If this part of memory can be shared, there will be a significant decrease in per-core memory %usage. Therefore, a multi-threaded simulation is necessary. 

To maximize the CPU resource utilization efficiency during massive MC production, we need to ensure that memory usage per CPU core matches the configuration of JUNO computing centers. The dominant part of memory usage is the physics list and detector geometry construction. If this part of the memory can be shared, there will be a significant decrease in per-core memory usage. Therefore, a multi-threaded simulation is necessary.

\section{Multi-threaded simulation framework}
\label{sec:multithread}

\subsection{Introduction to multi-threaded SNiPER framework}

To support multi-threading in the SNiPER framework, the MT-SNiPER extension had been implemented based on Intel Threading Building Blocks (TBB)~\cite{Zou:2018dqs, Zou:2019cyq}. Intel TBB is a flexible C++ library to simplify parallel programming in complex applications, supporting task-based programming. A module named Muster (Multiple SNiPER Task Scheduler) is developed to manage the execution of \texttt{Task}. As mentioned earlier, \texttt{Task} is a lightweight application manager. In sequential mode, \texttt{Algorithm}s and \texttt{Service}s are registered into a \texttt{Task} and then executed by the \texttt{Task}. Muster integrates both the \texttt{Task} component and TBB worker. Before starting the event loop, multiple instances of \texttt{Task} are created. These \texttt{Task} instances will not be executed immediately. Muster will schedule and execute them in the corresponding TBB workers. 

Data management and I/O modules are developed as well in the MT-SNiPER. Due to event correlations in physics analysis, the correct time order of events is important. The MT-SNiPER introduces I/O \texttt{Task}s and a thread-safe global buffer. An input \texttt{Task} is executed in an additional thread to read the events from files into memory. Each event is pushed to the global buffer. When the number of input events reaches a threshold, the input \texttt{Task} will be hold to avoid loading too many events. Then, the events in global buffer will be accessed by the \texttt{Task}s in SNiPER Muster. When an event is assigned to a \texttt{Task}, an internal flag will be marked to indicate that this event is under processing. When the \texttt{Task} finishes the processing of the event, the internal flag will be updated as done. Finally, an output \texttt{Task} will save these events into a file and remove them from the global buffer.

\subsection{Integration of MT-SNiPER and Geant4}

\begin{figure}[htb]\centering
\includegraphics[width=65mm]{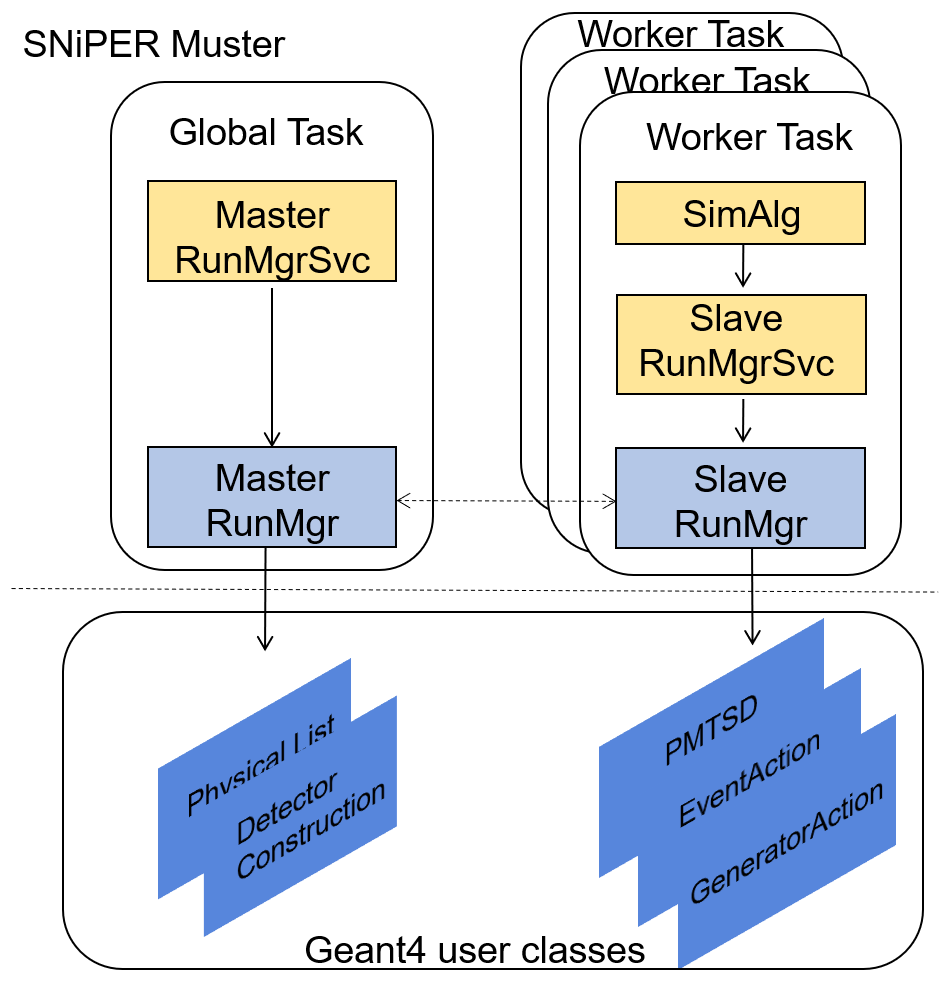}
\caption{Multi-threaded architecture of detector simulation}
\label{fig:mt_detsimstruct}
\end{figure}

\begin{figure}[htb]\centering
\includegraphics[width=85mm]{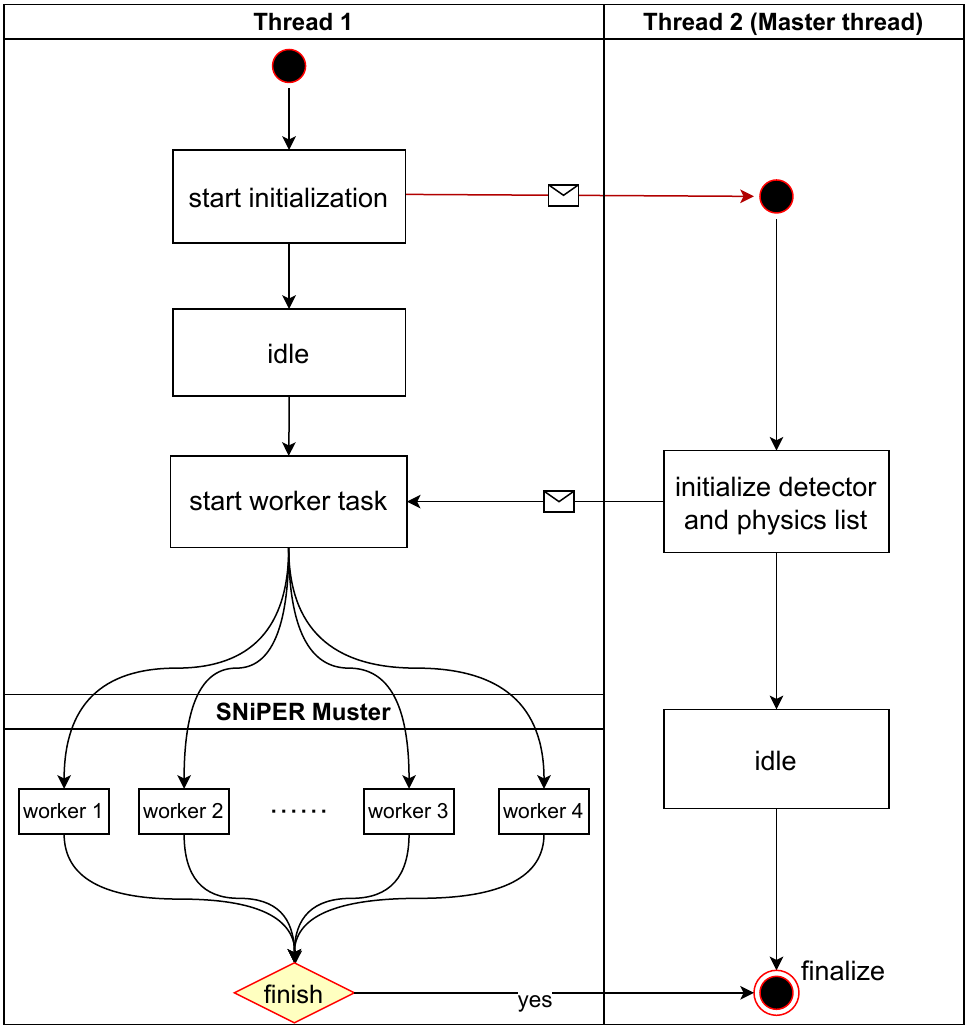}
\caption{Multi-threaded detector simulation construction process}
\label{fig:mtBuildProcess}
\end{figure}

According to the design of MT-SNiPER, the integration of Geant4 kernel should follow the same schema. However, the multi-threading in Geant4 is implemented in pool of threads. Paper~\cite{Dotti:2016ors} provides an example to integrate TBB and Geant4, introducing the task-based parallelism. Based on this example, an integration of MT-SNiPER and Geant4 is proposed~\cite{Lin_2018}. The architecture is shown in Figure \ref{fig:mt_detsimstruct} and the workflow is shown in Figure \ref{fig:mtBuildProcess}. First, a special SNiPER \texttt{Task} named Global Task is developed to manage the initialization of master run manager. This task is invoked before the Muster starts the event processing. As Geant4 master run manager needs to live in its own thread~\cite{Dotti:2016ors}, the Global Task spawns a new thread. The master run manager initializes the detector construction and physics list in Geant4, which are then used by the slave run manager in worker tasks. After the initialization, the Muster starts all the workers. The tasks in workers initialize the algorithm and the slave run manager at beginning. The slave run managers could access the master run manager and initialize the corresponding user actions. At the event loop of the tasks, the slave run managers are invoked to simulate events from the global buffer. Compared to the thread-based simulation, task-based simulation can benefit from dynamical execution . For example, in thread-based simulation, the number of events in each thread is fixed at beginning. If simulation of an event takes a long time, the rest of events in the same thread will be blocked. If using task-based simulation, then the rest events could be scheduled to the other threads. 

\subsection{Event generation in multi-threaded simulation}

\begin{figure}[htb]
    \centering
    \includegraphics[width=85mm]{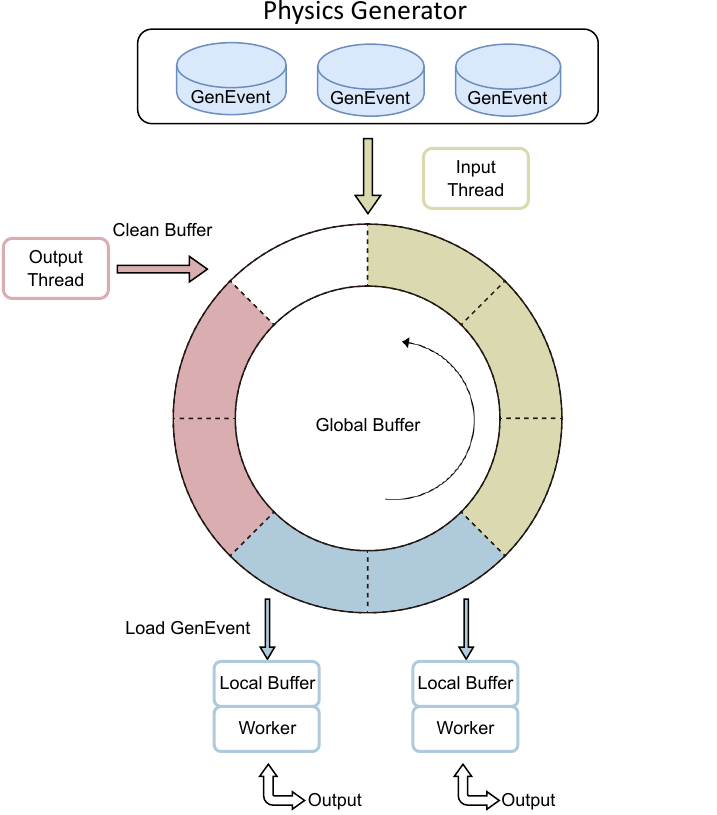}
    \caption{Operation mechanism of multi-threaded detector simulation}
    \label{fig:globalbuffer}
\end{figure}

Physics generators in JUNO mostly produce events into files in sequence. Since they are maintained as standalone projects, parallelizing these generators is challenging. In order to integrate them into the multi-threaded simulation framework, the generation of events is moved to the input \texttt{Task} of the MT-SNiPER. The data flow is shown in Figure \ref{fig:globalbuffer}. Physics generator produces \texttt{GenEvent} objects into its own buffer in the input \texttt{Task}. Then these events are moved to the global buffer by the underlying framework. The global buffer in the MT-SNiPER is implemented as a thread-safe circular buffer, which could be shared by workers. The workers retrieve the \texttt{GenEvent} objects and put them into the local buffers of the workers. Then, simulation algorithms in workers process these events. When the simulation of an event is done, the event is marked as done. The output \texttt{Task} is responsible to write the events into file in sequence and remove them from the global buffer. 

A problem during event generation is that the vertices of events need to be randomized, which requires the full detector geometry information in Geant4. As mentioned in the previous section, the Geant4 kernel is initialized in another thread, which is not available in the input \texttt{Task}. In order to randomize the vertices, this part is moved from the input \texttt{Task} to the workers. Instances of generator algorithm are created in the workers. The handling of \texttt{GenEvent} is different from the input \texttt{Task}. In the input \texttt{Task}, new \texttt{GenEvent} objects are created. In the workers, the \texttt{GenEvent} objects are not created but loaded from the global buffer. Then these objects are modified by the position generator tools. At this stage, the position generators can access the Geant4 geometry.

\subsection{Event reproducible and random seed}

Another issue is reproducible simulation between single-threaded and multi-threaded simulation. In single-thread detector simulation, random number engine is initialized at the beginning of program execution by setting an initial seed value for this job. This could keep the event reproducible if the conditions are same. However, in the current task-based simulation framework, the events are simulated dynamically, which is hard to reproduce the same results. For example, a job with same input could be configured with different CPU cores. The random number engines will be different with different CPU cores. Even though a job is configured with same CPU cores, the events could be assigned to different instances, which is different from the previous. 

When users need to reproduce the simulation results, the simulation framework provides two different ways, whose inputs are different. The first way is that all the events are re-generated from the beginning with the same input file from physics generators.

Random number engine initialized at the beginning of program execution can ensure that the \texttt{GenEvent} with the same event ID is consistent, such as vertices. Before simulating in a single thread, the random number sequence will be set based on the current event ID and user-specified initial seed value. Similarly, in multi-threading, we will reset the random number sequence before Worker retrieves the \texttt{GenEvent} object to start simulation to ensure consistency of simulation results in both single-threaded and multi-threaded mode.

\subsection{Unified user interface for serial and multi-threaded simulation}
Due to the different implementations for serial and multi-threaded simulations, the configurations are different. For example, an instance of the position generator tool is registered into the generator algorithm for the serial mode, while multiple instances of this tool are registered into the generator algorithms in workers for the multi-threaded mode. Such differences will cause confusion for the end users. Two Python modules are used to mask the difference. One module is for serial mode, while another is for multi-threaded mode. The latter is derived from the former. An option \texttt{nthreads} is used to control which module will be loaded. When this option is not specified, the module for serial mode is used by default. If the option is specified with number of workers, then the module for multi-threaded mode is used. By using these modules, users don't need to worry about of the differences. For the simulation software developers, most of the changes are in the module for serial mode. Therefore, there is little impact on them.

\section{Performance}
\label{sec:performance}

\begin{figure}[ht]
    \centering
    \includegraphics[width=85mm]{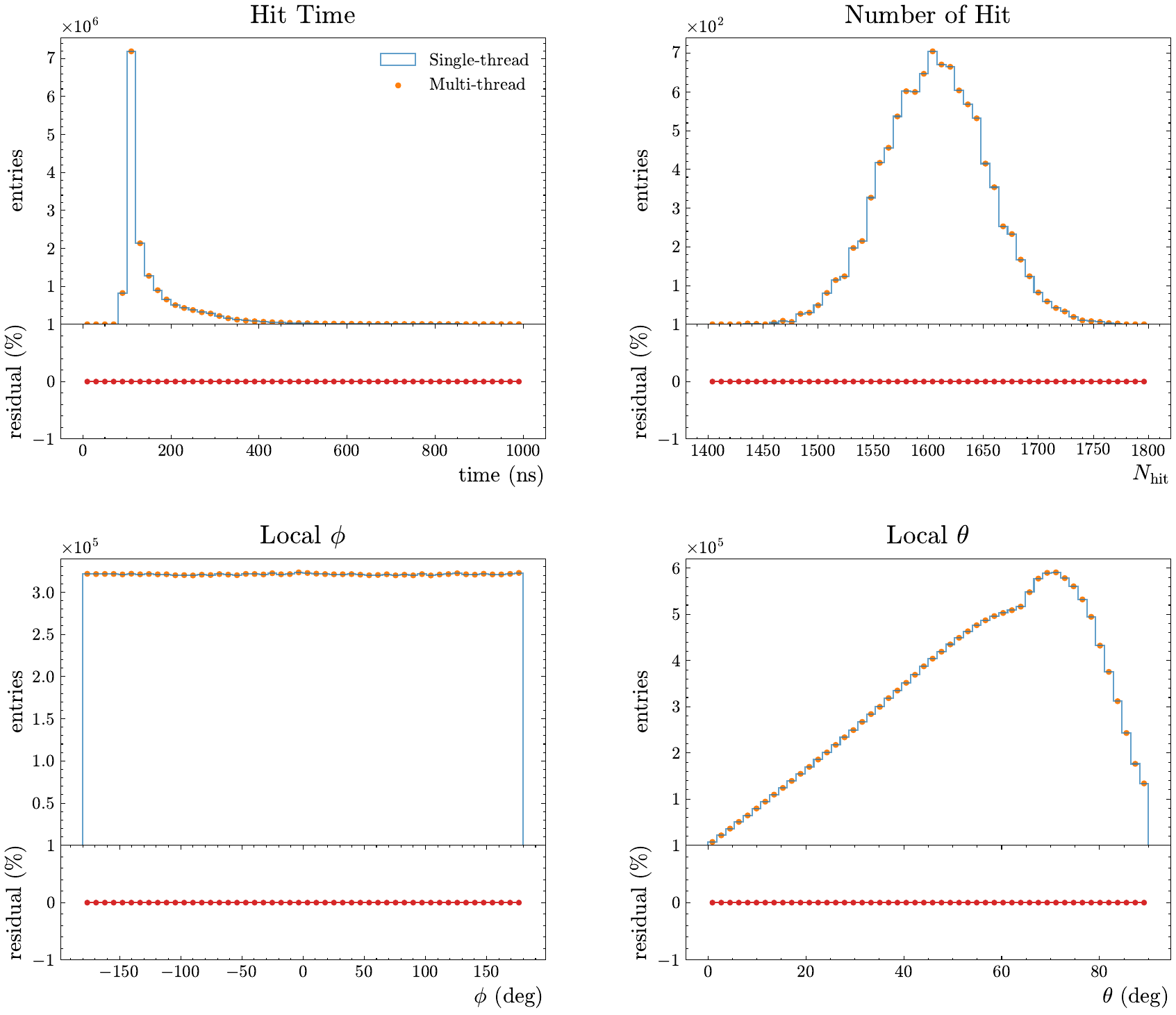}
    \caption{Results of multi-threading and single-threading detector simulation. The distributions of hit time, number of hits, as well as the $\phi$ and $\theta$ (the hit position in PMT local coordinates) were compared separately.}
    \label{fig:outputdata}
\end{figure}%

\begin{figure*}[ht]
    \centering
    \begin{minipage}{\textwidth}
        \centering
        \includegraphics[width=1\textwidth]{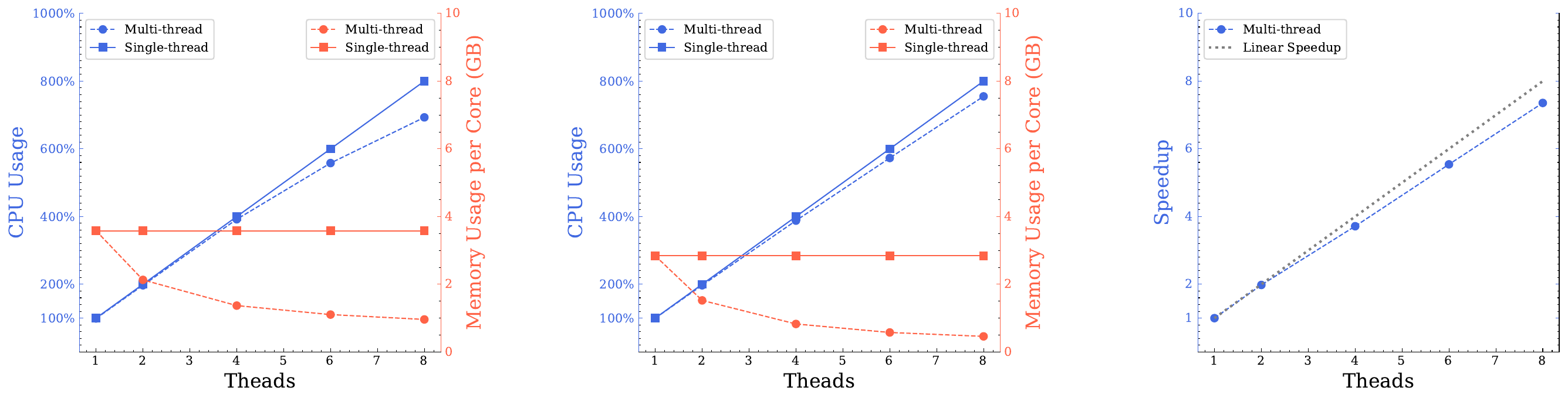} % 
        \caption{Multi-threading simulation performance (The left figure shows the performance test results of 200 muon events per job, while the middle figure shows the performance test results of 500k 1~MeV gamma-ray events. The CPU Usage percentage determined by calculating the sum of User Time and System Time, and dividing it by the Total Elapsed Time). The right figure illustrates the speedup achieved through multi-threading for the same job of 200 muon events.}
        \label{fig:performance}
    \end{minipage}
\end{figure*}%

\begin{figure}[ht]
    \centering
    \includegraphics[width=75mm]{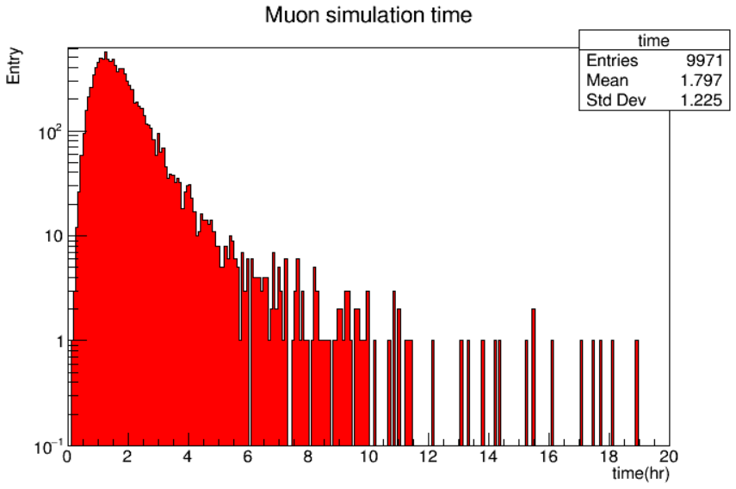}
    \caption{CPU time of muon event detector simulation.}
    \label{fig:MuonCPUTTime}
\end{figure}%

In the performance testing of the software, multi-threaded simulation tests for large memory computing jobs were conducted. Tests were carried out on same local clusters, with different kinds of physics events. 
% The detailed configurations of local clusters are as follows:

First, it is necessary to ensure the consistency of results between single-threaded and multi-threaded simulations. Due to the new random seed management system, consistent generator information and random number sequences are ensured in both single-threaded and multi-threaded simulations. Therefore, we conducted tests on the results of gamma-ray events from detector simulations, as shown in Figure \ref{fig:outputdata}, which demonstrate consistent results across the simulations whether they are run on single or multiple threads.

The performance test of 200 muon events was organized while ensuring the consistency of the results. Figure \ref{fig:performance} (the left one) shows that almost 100\% CPU utilization can be achieved with 2 threads. As the number of threads continues to increase, CPU usage will decrease. Figure \ref{fig:MuonCPUTTime} shows the CPU time of muon event simulation. The CPU time for muon event simulation varies from seconds to about 20 hours, which is due to the number of muon tracks, the muon energy, the muon track length and energy deposition in the liquid scintillator being different event by event. The number of optical photons generated during the muon passing through the detector determines the CPU time for each event. The large fluctuation of CPU time leads to the decrease in CPU usage. The reason for this phenomenon is that at the end of the simulation job, only one worker is simulating high-energy events while there are no other physical events available in the global buffer for simulation, resulting in other workers being idle. The program needs to wait for all events simulation to complete before entering the finalization stage. This results in a loss of CPU utilization in multi-threaded jobs. In the simulation of 500k 1~MeV gamma-ray events, simulation time is relatively even without multiple workers waiting. The CPU utilization in multi-threaded mode shows a linear improvement with minimal CPU usage loss, as shown in Figure \ref{fig:performance} (the middle one).

\begin{figure}[!ht]
    \centering
    \includegraphics[height=95mm]{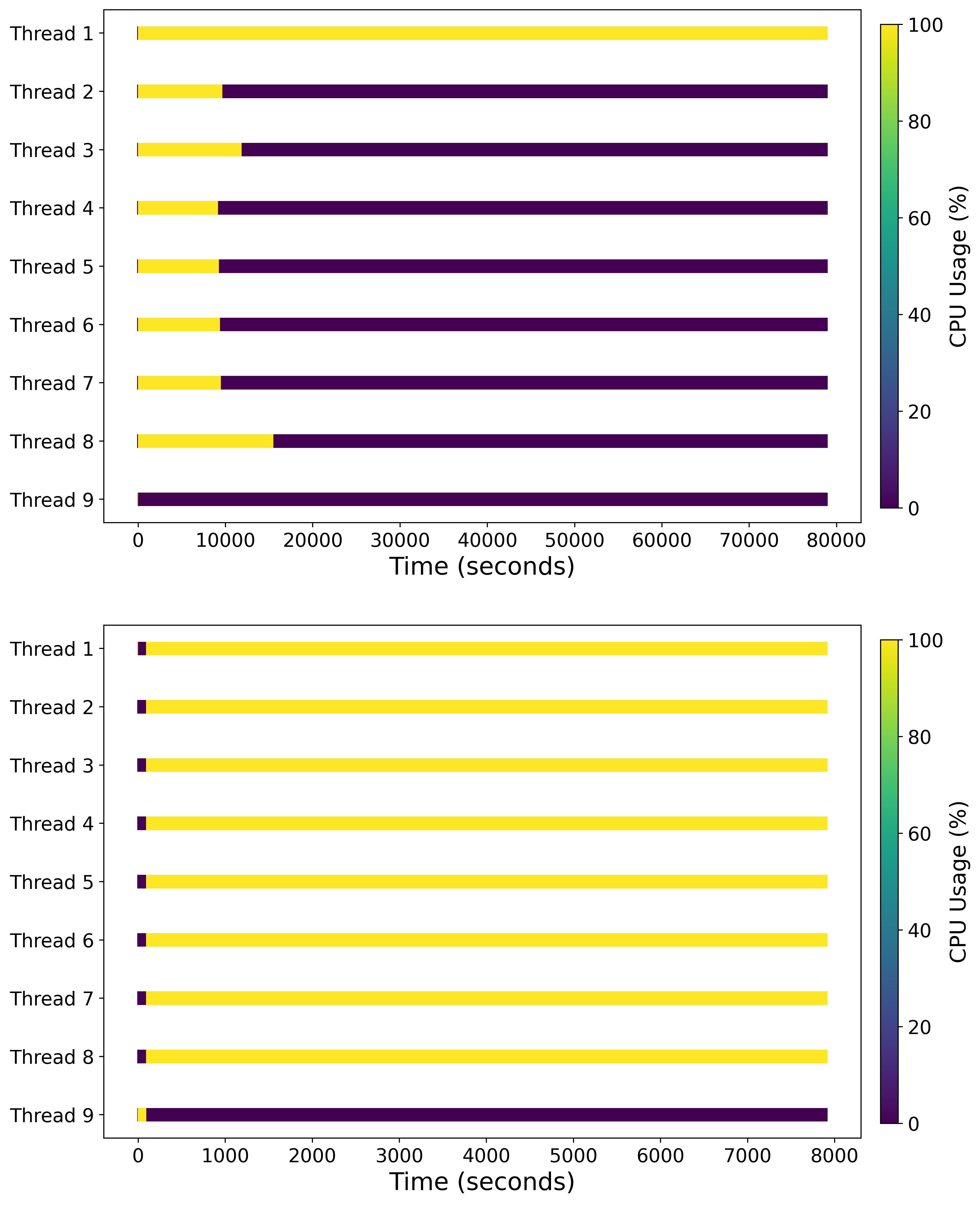}
    \caption{CPU utilization over time in eight-threaded jobs. The top graph shows a typical example of efficiency decrease in multi-threading in muon simulation, where other threads are waiting for the completion of one high-energy event simulation. The bottom graph shows the case of uniform distribution of simulation time in Gamma-ray simulation, resulting in higher CPU utilization. Thread~9 is the master thread used to complete program initialization in the global task.}
    \label{fig:cpusage_overtime}
\end{figure}

 Figure \ref{fig:cpusage_overtime} also confirms this point. We tracked the CPU utilization of each thread throughout the process of running multi-threaded jobs. In a simulation job with low CPU utilization for muon events, it can be seen that other threads are waiting for a high-energy event simulation to finish, causing overall low CPU utilization for the entire job. In gamma-ray event testing, where event simulation time is uniform and each thread runs at full load until program completion, higher CPU utilization is achieved.

Regarding memory usage in multi-threaded jobs, after starting multiple threads and benefiting from shared memory space required by detector geometry and physical processes as well as applying various memory reduction methods, memory per core has been reduced significantly, for example less than 2GB for 4 threads case, which fits well with computing center's job configurations without wasting computational resources.

For jobs requiring more threads, the optical simulation of high-energy events can be very time-consuming on the CPU. It is possible to select the job involving high-energy deposition events in liquid scintillation not to be simulated on the CPU. Opticks~\cite{Blyth:2019yrd} is an open-source software based on NVIDIA's OptiX ray tracing technology, designed to simulate the transport of photons on GPU. The JUNO detector simulation software utilizes Opticks to handle the transport of photons, while other components continue to be managed by the traditional Geant4 software. This heterogeneous computing approach allows the parallelizable parts of the simulation to be processed on the GPU, leveraging its powerful parallel computing capabilities to accelerate the simulation. In this way, it addresses the performance bottleneck of multithreading on the CPU and avoids wasting GPU computing resources on simulating low-energy deposition events.

\section{Summary}
\label{sec:summary}
In this paper, we propose a method of multi-threaded detector simulation based on JUNOSW and Geant4 frameworks to address the memory challenges brought by muons and atmospheric neutrinos in detector simulation, preparing for the generation of large amount of data needed by JUNO experiment.

To solve the memory issues caused by high-energy events, our development work is divided into two parts. One is the development of the DataModelWriterWithSplit method to solve existing memory problems in single-threading. The other is the development of a multi-threaded detector simulation architecture that ensures thread safety and synchronization among threads. In multi-threaded detector simulation, detector geometry and physics lists only need to be initialized once and can be shared by all threads. The DataModelWriterWithSplit method is used for outputting. These two aspects significantly reduce memory usage in both detector simulation and data I/O, achieving the target below 3GB per core and improving CPU utilization efficiency for overall data production.

Furthermore, after data generation and performance testing, we found that in terms of result comparison, due to the introduction of random engine design in the new detector simulation, both single-threaded and multi-threaded simulations can achieve identical results for jobs with identical event ID values and user-specified seed values. In the aspect of performance testing, we perform on local clusters; it is found that good test results could be achieved when using 4 threads in actual data production. For jobs with more threads, one main limitation comes from large fluctuation of energy deposition in liquid scintillator event by event, which cause large fluctuation of CPU time per event. Towards the end stage of program execution leading to idle threads. Future work should offload high energy deposition events to GPU, aiming to maximize CPU utilization potential.

\acknowledgments
We gratefully acknowledge support from National Natural Science Foundation of China (NSFC) under grant No. 12275293, 12375195 and 12025502. This work is also supported in part by  the Strategic Priority Research Program of the Chinese Academy of Sciences (CAS), Grant No. XDA10010900, supported in part by Youth Innovation Promotion Association of CAS under grant No. 2022011.

% Bibliography

%% [A] Recommended: using JHEP.bst file
%% \bibliographystyle{JHEP}
%% \bibliography{biblio.bib}

%% or
%% [B] Manual formatting (see below)
%% (i) We suggest to always provide author, title and journal data or doi:
%% in short all the informations that clearly identify a document.
%% (ii) please avoid comments such as "For a review'', "For some examples",
%% "and references therein" or move them in the text. In general, please leave only references in the bibliography and move all
%% accessory text in footnotes.
%% (iii) Also, please have only one work for each \bibitem.

\bibliographystyle{JHEP}
\bibliography{biblio.bib}

\end{document}